\documentclass[11pt,letterpaper]{article}
\usepackage{systeme} 
\usepackage[utf8]{inputenc}
\usepackage{multirow} 
\usepackage{booktabs} 
\usepackage[english]{babel}
\usepackage{amsmath}
\usepackage{amsfonts}
\usepackage{amssymb}
\usepackage{graphicx}
\usepackage[small,bf]{caption} 
\usepackage[left=3cm,right=3cm,top=2cm,bottom=3cm]{geometry}
\usepackage{framed} 
\usepackage{color} 
\usepackage{wrapfig}\definecolor{shadecolor}{RGB}{220,220,220} 
\usepackage{float} 
\usepackage{array} 
\usepackage{caption} 

\author{Jorge Pinochet}
\title{\textbf{Networked Observatory for Virtual Astronomy (NOVA): Teaching astronomy with AI}}
\begin{document}

\author{Jorge Pinochet$^{*}$\\ \\
 \small{$^{*}$\textit{Facultad de Ciencias Básicas, Departamento de Física. }}\\
  \small{\textit{Centro de Desarrollo de Investigación CEDI-UMCE,}}\\
 \small{\textit{Universidad Metropolitana de Ciencias de la Educación,}}\\
 \small{\textit{Av. José Pedro Alessandri 774, Ñuñoa, Santiago, Chile.}}\\
 \small{e-mail: jorge.pinochet@umce.cl}\\}

\date{}
\maketitle

\begin{center}\rule{0.9\textwidth}{0.1mm} \end{center}
\begin{abstract}
\noindent The term “artificial intelligence” (AI) first appeared in the mid-20th century. It wasn't until well into the 21st century that AI began to flourish. Today, AI has found very diverse applications. One area that is just beginning to be explored is physics education. The aim of this work is to contribute to this burgeoning educational field through a ChatGPT application named Networked Observatory for Virtual Astronomy (NOVA). This application provides students with simulated astronomical data in a ChatGPT-based environment, where they can put their scientific skills to work by formulating hypotheses, analyzing data, and drawing verifiable conclusions through interaction with ChatGPT.\\ \\


\begin{center}\rule{0.9\textwidth}{0.1mm} \end{center}
\end{abstract}

\maketitle

\section{Introduction}
The modern origins of the concept of artificial intelligence (AI) date back to the pioneering work of Allan Turing, John McCarthy, Marvin Minsky, Claude Shannon, and Allen Newel in the mid-20th century. However, it was not until well into the 21st century that AI began to flourish. Today, AI is one of the most active and transformative areas of science and technology, and has found applications in fields as diverse as medicine, finance, engineering, and software development, among others [1–3]. In recent years, AI has become widespread, becoming an everyday tool for millions of people around the world, thanks to the emergence of AI-assistants, also known as chatbots, such as ChatGPT or DeepSeek. These are AI-powered computer programs that simulate human conversation through text or voice, and that handle enormous volumes of information, offering solutions to a wide variety of queries and tasks.\\

AI is also finding innovative educational applications [4–6], but this is a field that is just beginning to be explored [7–15], so there is much work to be done. The objective of this work is to contribute to this burgeoning field of educational innovation through a ChatGPT application that I have called Networked Observatory for Virtual Astronomy (NOVA), which is intended to enhance the learning of physics and astronomy among secondary school students. NOVA provides students with simulated astronomical data in a ChatGPT-based environment, where they can put their scientific knowledge and skills to work by formulating hypotheses, analyzing data, and drawing verifiable conclusions through interaction with NOVA. This application of AI offers exciting learning opportunities, especially considering that collecting astronomical data often requires advanced technologies that are beyond the reach of most schools. Therefore, a virtual observatory like NOVA can help improve and democratize astronomy teaching at the secondary school level.\\

Although NOVA has diverse applications, here we will explore one of the most fundamental for astronomy education: verifying and applying Kepler's third law. To do so, we will use a scenario simulated by the ChatGPT, where a group of planets orbit a fictional star. In the first part of this article, we will introduce the theoretical framework, analyzing Kepler's third law. We will then describe the NOVA-based activity in detail. The article concludes with some comments and suggestions for maximizing educational benefits from NOVA.

\section{Theoretical framework: Kepler's third law}
Let us consider a massive spherical celestial body of mass $M$, such as a star, around which revolves a planet of mass $m\ll M$, describing an orbit that, for simplicity, we will assume is circumferential and has radius $R$. According to Newton's law of gravitation, the magnitude of the force $F$ exerted by the star on the planet is calculated as [16]:

\begin{equation}
F= \frac{GMm}{R^{2}},
\end{equation}

where $G = 6.67\times 10^{-11} N\cdot m^{2}\cdot kg^{-2}$ is the gravitational constant, and $R$ is the distance between the center of the planet and the center of the star (Fig. 1). The force of gravity experienced by the planet provides the centripetal force $mv^{2}/R$ that keeps it in a circumferential orbit with speed $v$. Taking $F=mv^{2}/R$ in Eq. (1) we get

\begin{equation}
v^{2} =\frac{GM}{R}.
\end{equation}

\begin{figure}[h]
  \centering
    \includegraphics[width=0.3\textwidth]{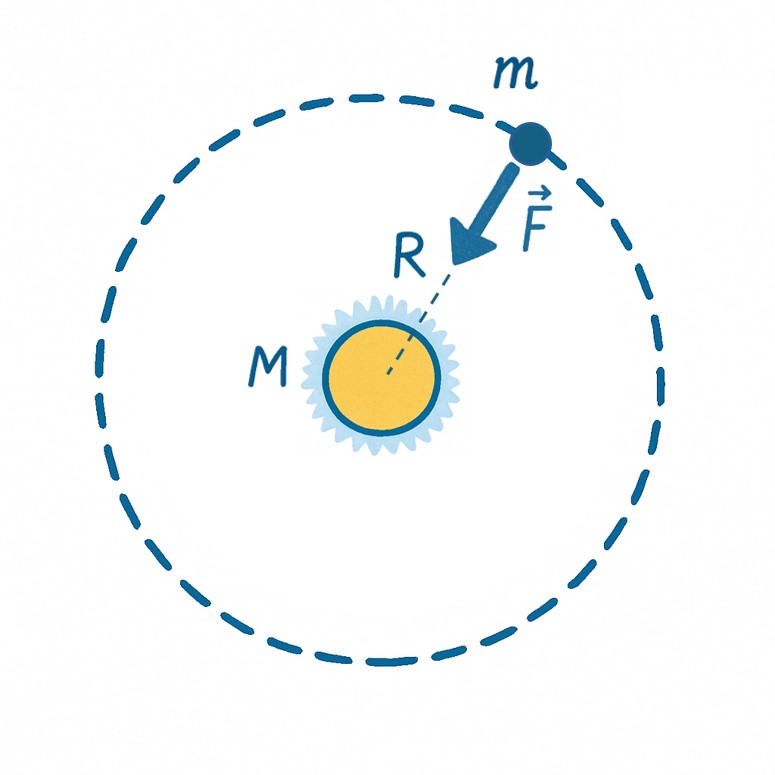}
  \caption{A planet of mass $m$ describes a circular orbit of radius $R$ around a star of mass $M\gg m$.}
\end{figure}

Considering that the speed of the planet is constant and is given by $v=2\pi R/T$, where $T$ is the period of revolution around the star, from Eq. (2) it results

\begin{equation}
\left( \frac{2\pi R}{T} \right)^{2} =\frac{GM}{R}.
\end{equation}

After a bit of algebra we obtain Kepler's third law [17]

\begin{equation}
\frac{R^{3}}{T^{2}} = \frac{GM}{4\pi^{2}} =K,
\end{equation}

where $K=GM/4\pi^{2}$  is a constant for each planetary system considered. It is important to emphasize that this result is only valid when the mass of the central object is much greater than that of the bodies orbiting it, as occurs when a group of planets revolves around a star.

If we plot Kepler's law, placing $T^{2}$ on the horizontal axis and $R^{3}$ on the vertical axis, we obtain a line like the one illustrated in Fig. 2, where the slope $K$ is proportional to the mass of the star. Consequently, knowing the planet's period of revolution $T$ and the radius $R$ of its orbit, the last equation allows us to determine the star's mass using the relation [17]

\begin{equation}
M=\left( \frac{4\pi^{2}}{G} \right) \frac{R^{3}}{T^{2}} = \left( \frac{4\pi^{2}}{G} \right)K.
\end{equation}

For the solar system, if we express $T$ in years and $R$ in astronomical units ($AU$), where $1AU = 1.5\times 10^{11} m$ is the mean distance between the Earth and the Sun, then $K=1AU^{3}\cdot year^{-2}$, and the equation $R^{3}$ vs $T^{2}$ will be a line inclined at $45^{o}$. If we plot Eq. (5) for any planetary system in the same units, any change in the slope with respect to $45^{o}$ indicates that the mass of the star is different from that of the Sun. This gives us another method to calculate the stellar mass.\\

\begin{figure}[h]
  \centering
    \includegraphics[width=0.3\textwidth]{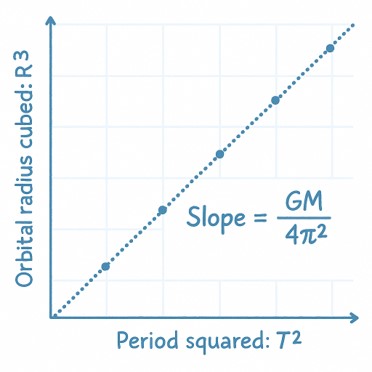}
  \caption{The graph $R^{3}$ vs $T^{2}$ corresponds to a line with a slope proportional to $M$.}
\end{figure}

Indeed, let us write Eq. (5) in terms of $K$, with the radius expressed in $AU$ and the period in years. For any planetary system this equation is written as $M=(4\pi^{2}/G)K$, while for the solar system it is written as $M_{\odot}=(4\pi^{2}/G)$ where $K=1$ and $M_{\odot}=1.99\times 10^{30} kg$ is the mass of the Sun. Under these conditions, if we divide both equations we obtain that

\begin{equation}
M=KM_{\odot}.
\end{equation}

Knowing the slope of a $R^{3}$ vs $T^{2}$ plot, we can calculate $M$ in solar mass units and in SI units from the slope of the equation of the corresponding line. In the next section, we will develop an activity where students will be tasked with verifying the validity of Kepler's third law and then using it to determine the mass of a star belonging to a fictional planetary system.

\section{Networked observatory for virtual astronomy (NOVA)}
The activity consists of seven stages (Fig. 3) and uses ChatGPT as the AI-assistant. The reason for this choice is that this chatbot has graphical resources and computing capabilities that are essential for the development of the activity. For example, the DeepSeek AI-assistant does not have graphical resources, although its computing capabilities are comparable to those of ChatGPT. Therefore, the activity requires that the students or their teacher have a ChatGPT account. The activity works with both the free version of ChatGPT and the Plus version.\\

There are two options for developing the activity. The first is to use a projector or a virtual whiteboard, so that all students have access to a single account and therefore a single simulation. The second option is for each student or group of students to use their own computer and therefore work with their own simulation. In my experience, this second option works best because it provides greater richness and diversity of results to the activity.\\

Regardless of the option chosen, the seven stages of the activity maintain their basic structure. In Stage 1, the teacher introduces the activity; in Stage 2, students obtain simulated data using NOVA; in Stages 3, 4, and 5, they address various problems related to the simulation; in Stage 6, they must return to NOVA to compare their own results with the solutions to the posed problems; and in Stage 7, the activity concludes with a plenary discussion.\\ 

\begin{figure}[h]
  \centering
    \includegraphics[width=0.5\textwidth]{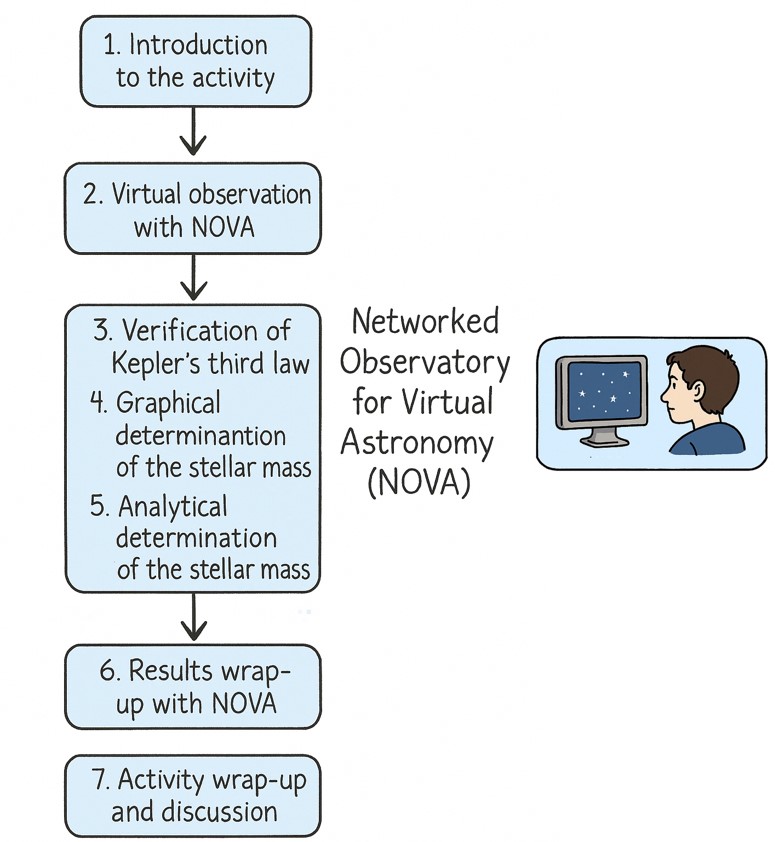}
  \caption{General structure of the activity.}
\end{figure}

As is customary in AI applications, we will use the word ``prompt" to designate the instructions given to the chatbot, and the word ``output" to designate its responses. The activity is described in detail below. To facilitate teachers' work, the activity has been written in Student Worksheet format, and the prompts and outputs presented correspond to a real interaction with ChatGPT during a typical activity. The activity assumes that students are familiar with using spreadsheets such as Excel.\\ 

To support teachers in developing this activity, two supplementary materials have been included in the appendix. The first is a guide on using NOVA with ChatGPT, and the second is a rubric for assessing student work. Both the guide and the rubric were generated using ChatGPT and then adapted by the author.\\

\textbf{Step 1: Introduction to the activity}\\
In this activity, you will work with NOVA (Networked Observatory for Virtual Astronomy), a virtual astronomical observatory based on ChatGPT, which simulates fictional planetary systems. Follow the teacher's instructions, who will provide you with the theoretical framework and form collaborative work groups. The activity requires working with six astronomical parameters shown in the Table 1.

\begin{table}[htbp] 
\begin{center}
\caption{Parameters involved in the activity}
\resizebox{0.7\textwidth}{!} {
\begin{tabular}{l l l} 
\toprule
\textbf{Parameter} & \textbf{Description} \\
\midrule

$T$ & Revolution period of a planet  \\ 

$R$ & Radius of the planetary orbit  \\

$M$ & Mass of the central star \\ 

$M_{\odot}$ & Solar Mass ($1.99\times 10^{30}kg$)  \\ 

$G$ & Gravitational constant ($6.67 \times 10^{-11} N\cdot m^{2} \cdot kg^{-2}$)  \\

$AU$ & Astronomical unit ($1.5\times 10^{11} m$)  \\

\midrule
\end{tabular}
}
\label{Parameters involved in the activity}
\end{center}
\end{table}

The objectives of the activity are:

{\small
\begin{itemize}
\item \textit{Apply and verify Kepler's third law in circular orbits}.\
\item \textit{Plot quadratic and cubic relationships from simulated experimental data}.\
\item \textit{Determine the mass of a star using Kepler's third law}.\
\item \textit{Evaluate the use of simulated models for the study of astronomical phenomena}.\
\end{itemize}
}

\textbf{Step 2: Virtual observation using NOVA}\\
To start working and have the astronomical simulation that allows you to verify and apply Kepler's third law, enter the following command to put the ChatGPT in NOVA mode:\\

\begin{figure}[H]
    \includegraphics[width=1\textwidth]{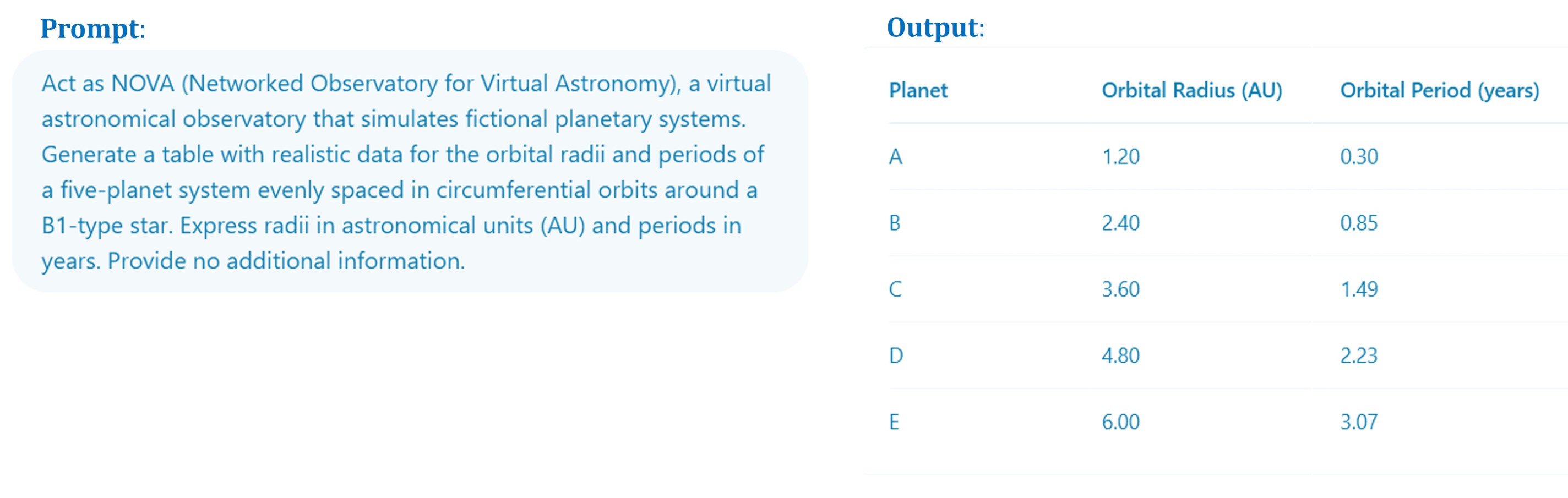}
\end{figure}

\textbf{Step 3: Verifying Kepler's third law}\\
Add two columns to the right of the table from step 2, one for $R^{3}$ and one for $T^{2}$. Using the data in these new columns, generate a plot with $R^{3}$ on the vertical axis and $T^{2}$ on the horizontal axis, using a spreadsheet.\\

\emph{Questions for group discussion}:
{\small
\begin{itemize}
\item Why is it necessary to generate these new columns?
\item What type of $R^{3}$ vs $T^{2}$ plot do you expect if Kepler's third law is correct?
\item What does the slope of the $R^{3}$ vs $T^{2}$ plot represent?
\end{itemize}
}

\textbf{Step 4: Graphical determination of stellar mass}\\
Plot the data from the table in Step 2 using a spreadsheet and generate the equation of the graph, known as the \textit{best fit line}. This equation provides the slope $K$ value, which will allow you to calculate the star's mass using the relationship:
\[M=KM_{\odot}.\]
Express your results in units of solar mass and in kilograms.\\

\emph{Questions for group discussion}:
{\small
\begin{itemize}
\item Why does the slope represent the mass of the star?
\item What would the slope be if it were the solar system?
\item What does a slope less than one represent?
\end{itemize}
}

\textbf{Step 5: Algebraic determination of stellar mass}\\
Convert the information in columns $R^{3}$ and $T^{2}$ from step 3 to SI units (seconds and meters). With this new information, calculate the star's mass based on the orbit of each of the five planets, using the equation
\[M=\left( \frac{4\pi^{2}}{G} \right) \frac{R^{3}}{T^{2}}.\]
Express your results in units of solar mass and in kilograms.\\

\emph{Questions for group discussion}:
{\small
\begin{itemize}
\item Do the results for the five planets agree?
\item Do the values of $M$ calculated in steps 4 and 5 agree?
\item Assuming there are discrepancies between steps 4 and 5, what could be the cause?
\item Are any discrepancies in the mantissas or in the powers of 10?
\end{itemize}
}

\textbf{Step 6: Confirming results with NOVA}\\
To find out if your answers to the problems posed in steps 3, 4 and 5 are correct, provide the following command to NOVA:

\begin{figure}[H]
    \includegraphics[width=1\textwidth]{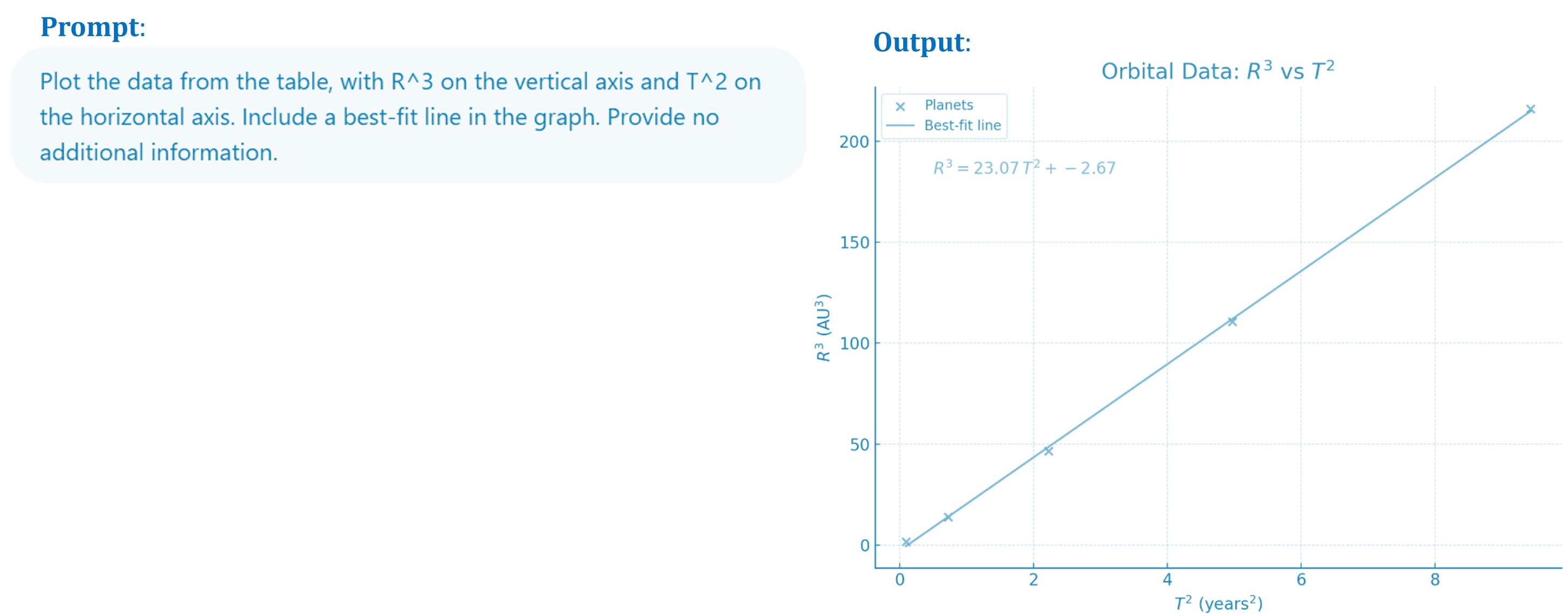}
\end{figure}

\begin{figure}[H]
    \includegraphics[width=1\textwidth]{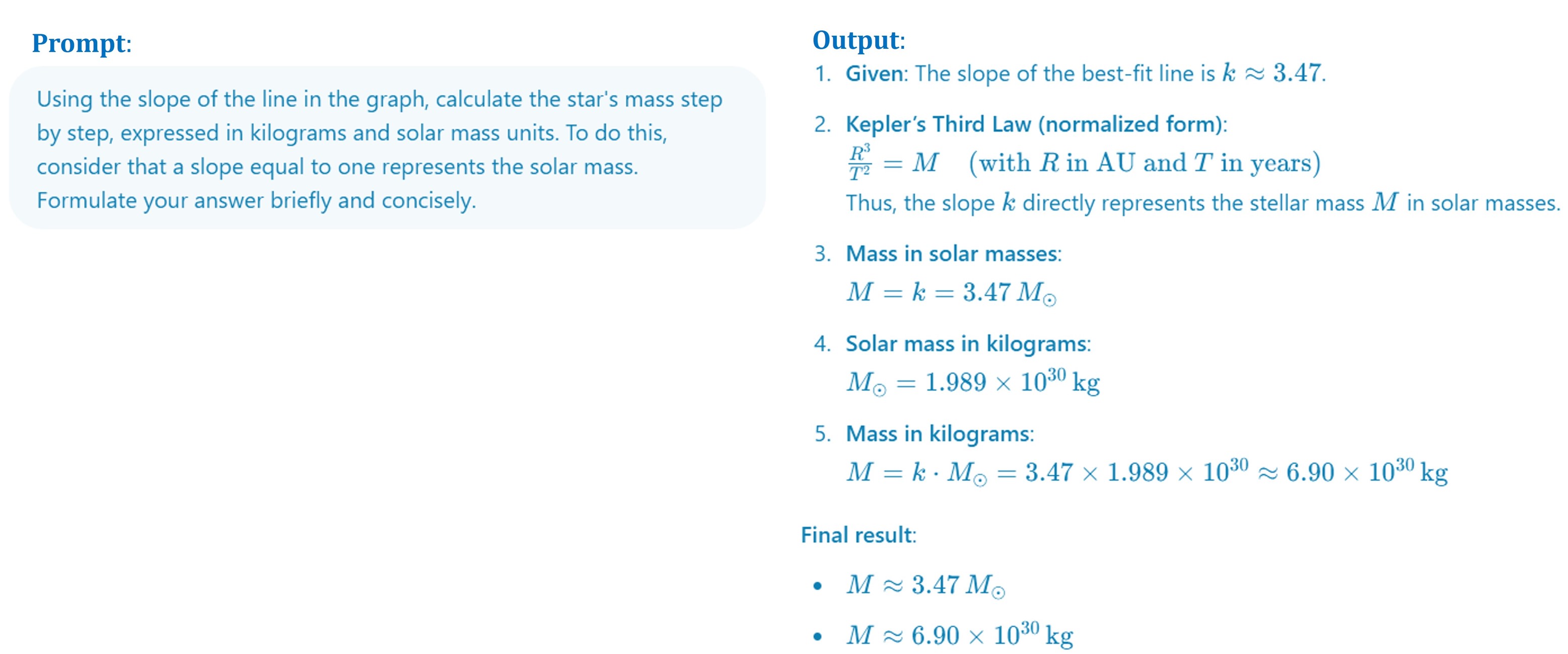}
\end{figure}

\begin{figure}[H]
    \includegraphics[width=1\textwidth]{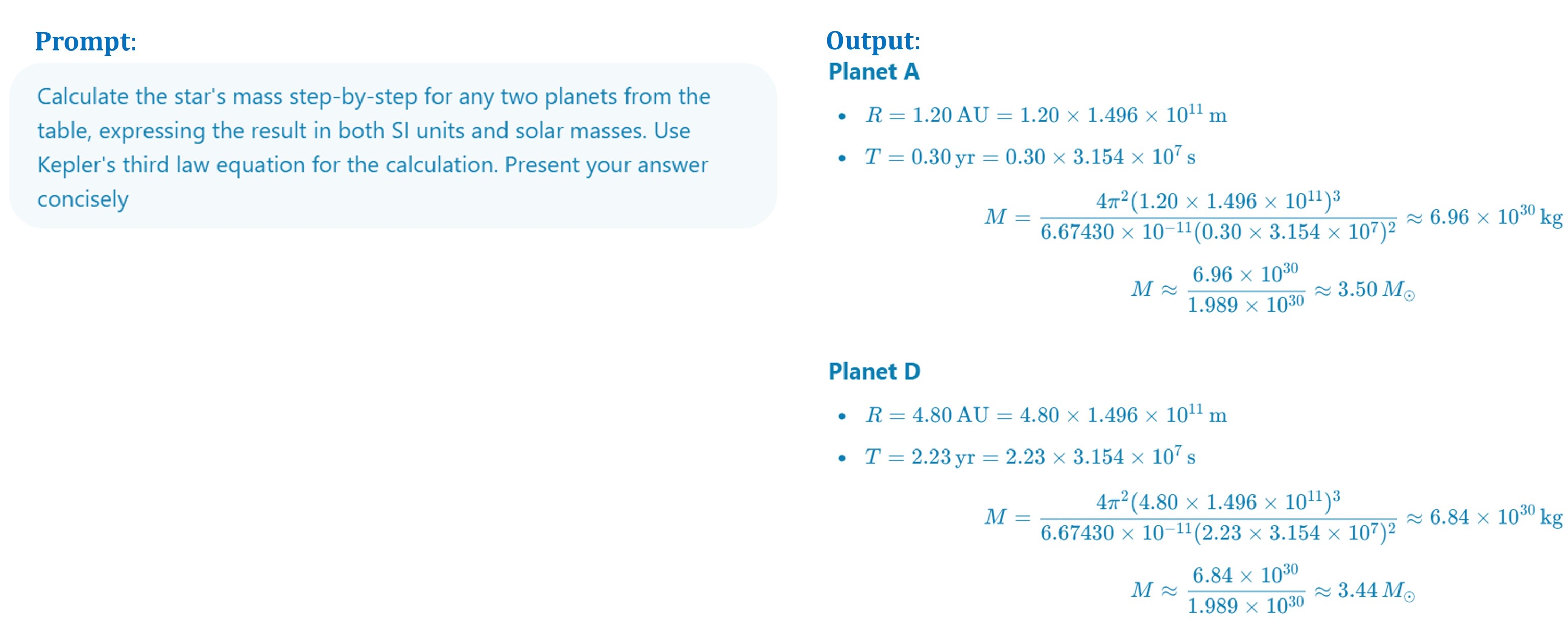}
\end{figure}

\emph{Questions for group discussion}:
{\small
\begin{itemize}
\item Does your graphical determination of the star's mass agree with the NOVA results?
\item Does your algebraic determination of the star's mass agree with the NOVA results?
\item If there is no agreement, where could you have made errors?
\end{itemize}
}

\textbf{Step 7: Closing the activity and plenary discussion}\\
Choose a representative from the group to show the results obtained to the teacher and the other students.\\

\emph{Questions to discuss with the group and the class}:
{\small
\begin{itemize}
\item What difficulties arose during the activity?
\item What were the main lessons learned from the activity?
\item What are the pros and cons of using simulated data in a virtual observatory?
\item Can you think of other applications of Kepler's third law using NOVA?
\item What other astronomy topics would you like to explore with NOVA?
\end{itemize}
}

\section{NOVA vs. traditional astronomy simulators}
To consolidate ideas and more easily highlight NOVA's features, it is useful to compare it with Stellarium, one of the most popular astronomy simulators. One of NOVA's most notable aspects is its flexibility, which allows it to generate fictional planetary systems with customized parameters, making it an ideal tool for adapting to specific educational objectives. In contrast, Stellarium simply delivers predefined, real-world astronomical data, with no modification options. Furthermore, as we have seen, NOVA promotes inquiry, allowing students to obtain data that they can tabulate, graph, and analyze, developing scientific and critical thinking skills, something that is difficult to achieve with Stellarium, which is essentially a visual resource.\\

Another notable aspect of NOVA is that it is a democratic educational tool, as its use only requires access to ChatGPT, making it suitable for use in schools with limited resources. Stellarium, on the other hand, requires installed software and powerful hardware, which considerably limits its use. Finally, one of NOVA's most innovative aspects, and what sets it apart from Stellarium and other astronomy simulators, is that it integrates astronomy, physics, and AI, allowing you to explore concepts like Kepler's laws with simulations that are impossible in reality.\\

Stellarium offers many visual resources and remains unsurpassed for realistic observations. Therefore, Stellarium could become a good complement to NOVA, offering possibilities and visual resources that are beyond the reach of an AI-assistant like ChatGPT.\\

\section{Concluding Remarks: Comments and suggestions for working with NOVA }

To guide teachers and maximize the benefits of NOVA, it is worth making a few comments regarding the ideas developed in the previous section.\\

To optimize time and make work more productive, it is advisable for students to divide up some tasks within the group, while maintaining teamwork. For example, if the group consists of three members, which is the maximum recommended number, there could be a prompt specialist, a data analyst, and a group representative who reports their results to the class.\\

The activity was tested and improved by the author through trial and error, requiring numerous iterations that led to a final version that was optimal for the students' work. During the process, the prompts were always short and concise, as this is a basic requirement to get the most out of a chatbot. Therefore, when students interact with NOVA, it is advisable that the prompts follow the structure presented in the previous section. Furthermore, since the activity is designed for high school students, all technical terms that are not strictly necessary have been avoided. Thus, concepts such as logarithmic plots, the eccentricity of planetary orbits, or the calculation of uncertainties have been intentionally avoided.\\

\begin{table}[H] 
\begin{center}
\caption{Possible activities with NOVA}
\resizebox{1\textwidth}{!} {
\begin{tabular}{m{7cm}  m{10cm} }
\toprule
\textbf{Discovery } & \textbf{Description} \\
\midrule

Inverse Square Law & Simulate the brightness of a star seen from different distances.   \\ 

Orbits of artificial satellites & Simulate satellites in circumferential orbits with different radii and speeds. \\

Light curves of transiting exoplanets & Simulate how the brightness of a star varies during the passage of a planet in front of it. \\ 

Kepler's first and second laws &  Simulate orbits with different eccentricities and observe how the speed changes at different points. \\ 

Cosmology and Hubble-Lemaître law & Simulate distance and recession velocity data for distant galaxies. \\

\midrule
\end{tabular}
}
\label{Possible activities with NOVA}
\end{center}
\end{table}

As we saw in the previous section, the first prompt asks NOVA for a simulation where the star is of B1 spectral type, a technical term that not all students are familiar with, but which they do not need to understand for the activity. The reason for introducing this technical term is to increase the accuracy of the chatbot's responses and thus improve the student experience. A type B1 star is a massive, hot, blue star belonging to the main sequence, which is the stage in a star's life where the energy source is the conversion of hydrogen into helium. Therefore, each teacher can feel free to choose the type of star that best suits them, but it is always advisable to specify the spectral type.\\

Although this may appear self-evident, it is necessary to emphasize that ChatGPT, like any other chatbot, is only a tool whose effectiveness will depend on how it is used. In the educational field, and particularly in the case of teaching astronomy through simulations, having the best chatbot is useless if the activity accompanying the simulations is not well designed or does not respond to the needs of teachers and their students. In this work, we have addressed one possible application of chatbots, but there is a wide range of possibilities that teachers can explore. Ultimately, the astronomical scenarios chosen for a simulation will depend on each school's curriculum, each teacher's preferences, and the students' interests. Kepler's laws are undoubtedly a fundamental topic, present in almost all secondary school physics curricula internationally, but teachers can also explore other applications. Table 2 contains five NOVA applications that only require simple calculations and graph interpretation.\\

Clearly, there are many other possibilities for physics teachers to explore, as the learning opportunities offered by AI are enormous. Making the most of this powerful tool -which seems destined to revolutionize the way we teach physics and astronomy- depends only on our motivation.

\section*{Appendix: Supplementary materials}

\emph{Teacher Guide: Using NOVA with ChatGPT in the Classroom}\\ 
To support teachers in implementing the NOVA (Networked Observatory for Virtual Astronomy) activity, anticipating common technical or pedagogical issues, and providing practical solutions.

\begin{table}[H] 
\begin{center}
\caption*{Common Issues and Solutions}
\resizebox{1\textwidth}{!} {
\begin{tabular}{ m{5cm}  m{5cm}  m{10cm} }
\toprule
\textbf{Issue} & \textbf{Possible Cause} & \textbf{Recommended Solution} \\
\midrule

ChatGPT does not generate the requested table & Prompt is poorly formulated or too vague & Rephrase the prompt with more precision. Example: “Generate a table with 5 planets orbiting a B1-type star, indicating $R$ in AU and $T$ in years.” \\ 

The $R^{3}$ vs $T^{2}$ plot is not a straight line & Errors in calculations or incorrect units & Check that students correctly calculated $R^{3}$ and $T^{2}$. Recalculate or compare with example values. \\ 

Inconsistent results between graphical and algebraic methods & Unit conversion errors or excessive rounding & Ensure students convert correctly to SI units (meters and seconds). Emphasize proper use of powers of 10. \\ 

The chatbot gives unhelpful or inaccurate responses & Prompt is too long, ambiguous, or informal & Keep prompts short, direct, and specific. Example: “What is the mass of the star if the slope is 0.85?” \\ 

Connectivity issues or website blocked & Unstable connection or school restrictions & Have a backup PDF with pre-generated simulated tables to work offline. \\ 

Students don’t understand concepts like 'slope' or 'proportionality' & Lack of prior conceptual understanding & Pause and briefly review these concepts with visual examples before continuing. \\ 

Students can't create ChatGPT accounts & Minimum age or institutional email required & Use a single teacher account projected to the class or group students to work collaboratively. \\ 

\midrule
\end{tabular}
}
\label{Common Issues and Solutions}
\end{center}
\end{table}

\emph{Assessment Rubric: NOVA Activity}\\  
This rubric is designed to evaluate students' performance during the NOVA (Networked Observatory for Virtual Astronomy) activity. It covers scientific understanding, data handling, teamwork, and communication.

\begin{table}[H] 
\begin{center}
\caption*{Assessment Rubric}
\resizebox{1\textwidth}{!} {
\begin{tabular}{ m{4cm}  m{4cm}  m{4cm} m{4cm}  m{4cm} }
\toprule
\textbf{Criterion} & \textbf{Excellent (4 pts)} & \textbf{Good (3 pts)} & \textbf{Satisfactory (2 pts)} & \textbf{Needs Improvement (1 pt)} \\
\midrule

Understanding of theoretical background	& Clearly explains Kepler's third law and its application. & Explains the law with minor omissions.	 & Shows partial or confused understanding. & Does not demonstrate understanding of the law. \\ 

Handling of simulated data	& Uses correct units, converts magnitudes, and performs calculations accurately. &  Minor or isolated errors in calculations. & Several errors affecting results. & Significant difficulty handling data. \\ 

Use of chatbot (NOVA)	& Formulates clear and effective prompts and uses responses well. & Prompts are somewhat ambiguous or need minor adjustments.	 & Needs considerable help to formulate or interpret prompts. & Unable to interact effectively with NOVA. \\ 

Interpretation of graphical results & Accurately interprets the graph and the slope. & Good interpretation with minor mistakes.	 & Limited understanding of the graph. &  Does not understand the graphical meaning.\\ 

Teamwork and collaboration	& Actively participates, cooperates, and communicates ideas clearly. & Participates but with less initiative. & Minimal participation or integration difficulties. &  No collaboration or participation.\\ 

Scientific communication and argumentation & Presents results with clear, logical, and well-justified language.	 & Communicates adequately with minor clarity issues.	 & Weak arguments or unclear communication.	 & Fails to communicate or justify results. \\ 

\midrule
\end{tabular}
}
\label{Common Issues and Solutions}
\end{center}
\end{table}

Maximum score: 24 points\\ 
Suggested grading scale:
\begin{itemize}
\item 22–24: Excellent performance
\item 18–21: Good
\item 14–17: Acceptable with improvement needed
\item <14: Requires additional support
\end{itemize}

\section*{Acknowledgments}
I would like to thank to Jorge Pinochet Jeric and Daniela Balieiro for their valuable comments in the writing of this paper. 

\section*{References}
[1] J. Černevičienė, A. Kabašinskas, Explainable artificial intelligence (XAI) in finance: a systematic literature review, Artif Intell Rev 57 (2024) 216. https://doi.org/10.1007/s10462-024-10854-8.

\vspace{2mm}

[2] Y. Hu, D. Yang, Z. Zhang, T. Yang, J. Peng, Application of Artificial Intelligence in Dynamic Image Recognition, J. Phys.: Conf. Ser. 1533 (2020) 032093. https://doi.org/10.1088/1742-6596/1533/3/032093.

\vspace{2mm}

[3] D.B. Olawade, A.C. David-Olawade, O.Z. Wada, A.J. Asaolu, T. Adereni, J. Ling, Artificial intelligence in healthcare delivery: Prospects and pitfalls, Journal of Medicine, Surgery, and Public Health 3 (2024) 100108. https://doi.org/10.1016/j.glmedi.2024.100108.

\vspace{2mm}

[4] M. Bond, H. Khosravi, M. De Laat, N. Bergdahl, V. Negrea, E. Oxley, P. Pham, S.W. Chong, G. Siemens, A meta systematic review of artificial intelligence in higher education: a call for increased ethics, collaboration, and rigour, International Journal of Educational Technology in Higher Education 21 (2024) 4. https://doi.org/10.1186/s41239-023-00436-z.

\vspace{2mm}

[5] S. Stankovski, G. Ostojić, S. Tegeltija, M. Stanojević, M. Babić, X. Zhang, Generative AI Applications and Tools in Engineering Education, in: 2024 23rd International Symposium INFOTEH-JAHORINA (INFOTEH), 2024: pp. 1–4. https://doi.org/10.1109/INFOTEH60418.2024.10495941.

\vspace{2mm}

[6] S. Wang, F. Wang, Z. Zhu, J. Wang, T. Tran, Z. Du, Artificial intelligence in education: A systematic literature review, Expert Systems with Applications 252 (2024) 124167. https://doi.org/10.1016/j.eswa.2024.124167.

\vspace{2mm}

[7] B. Billingsley, J.M. Heyes, T. Lesworth, M. Sarzi, Can a robot be a scientist? Developing students’ epistemic insight through a lesson exploring the role of human creativity in astronomy, Phys. Educ. 58 (2022) 015501. https://doi.org/10.1088/1361-6552/ac9d19.

\vspace{2mm}

[8] M.G. de Souza, M. Won, D. Treagust, A. Serrano, Visualising relativity: assessing high school students’ understanding of complex physics concepts through AI-generated images, Phys. Educ. 59 (2024) 025018. https://doi.org/10.1088/1361-6552/ad1e71.

\vspace{2mm}

[9] S. El-Adawy, I. Liao, V. Lad, M. Abdelhafez, P. Dourmashkin, Streamlining Physics Problem Generation to Support Physics Teachers in Using Generative Artificial Intelligence, The Physics Teacher 62 (2024) 595–598. https://doi.org/10.1119/5.0201458.

\vspace{2mm}

[10] B. Gregorcic, A.-M. Pendrill, ChatGPT and the frustrated Socrates, Phys. Educ. 58 (2023) 035021. https://doi.org/10.1088/1361-6552/acc299.

\vspace{2mm}

[11] R. Kahaleh, V. Lopez, Evaluating large language models in high school physics education: addressing misconceptions and fostering conceptual understanding, Phys. Educ. 60 (2025) 025013. https://doi.org/10.1088/1361-6552/adb235.

\vspace{2mm}

[12] T.O.B. Odden, A.S. Hansen, Physics Intuition About Vectors Can Help Us Understand Generative Artificial Intelligence, The Physics Teacher 63 (2025) 173–176. https://doi.org/10.1119/5.0149555.

\vspace{2mm}

[13] A.R. dos Santos Silva, G.L. Sales, NarraCubes: A Virtual Dice Simulator to Incorporate Storytelling in Physics Classes, The Physics Teacher 62 (2024) 678–681. https://doi.org/10.1119/5.0161738.

\vspace{2mm}

[14] A. Sperling, J. Lincoln, Artificial intelligence and high school physics, The Physics Teacher 62 (2024) 314–315. https://doi.org/10.1119/5.0202994.

\vspace{2mm}

[15] J.J. Trout, L. Winterbottom, Artificial intelligence and undergraduate physics education, Phys. Educ. 60 (2024) 015024. https://doi.org/10.1088/1361-6552/ad98de.

\vspace{2mm}

[16] W. Bauer, G. Westfall, University physics with modern physics, 1st ed., McGraw-Hill, New York, 2011.

\vspace{2mm}

[17] P.A. Tipler, Physics for Scientists and Engineers, W. H. Freeman and Company, New York, 2004.

\end{document}